\documentclass[12pt]{article}
\usepackage[dvips]{graphics}
\usepackage[dvips]{graphicx}
\begin{document}

\title{Surface Shape of Two-Dimensional Granular Piles}
\author{Bo Jin$^{\dag}$, $^{1,2}$  Fang Xu$^{\ddag}$ $^{1,2}$  and  Hern$\acute{{\mathrm a}}$n A. Makse$^{\S}$ $^{1,3}$\\
{\small {$^{1}$The Levich Institute}, {$^{2}$Department of Chemical Engineering}, {$^{3}$Physics Department}}\\
{\small {City College of New York}, {New York, NY 10031}}\\
{\small {$^\dag$ Email: jin$@$levdec.engr.ccny.cuny.edu}}\\
{\small {$^\ddag$ Email: fangxu$@$levdec.engr.ccny.cuny.edu}}\\
{\small {$^\S$ Email: makse$@$mailaps.org}}\\
}
\date{\today}
\maketitle
\begin{abstract}
\baselineskip 24pt
We study the surface
shape of two-dimensional piles using experiments and a
continuum theory for surface flows of granular materials (the BCRE equations).
We first obtain an analytical solution to the BCRE equations
with a simple transformation and show that the surface shapes thereby
predicted are in good agreement with the experimental results.
By means of such an analytical solution, we find that the formation
of the curved tails at the bottom of such piles depends not only on the properties of the granular
materials but also on the drift velocity of the grains within the rolling layer.
\end{abstract}

\newpage

\baselineskip 24pt

Granular materials have received increasing attention in recent years
on account of the fascinating granular flow patterns
and their important practical applications.
It is well-known that the angle of repose, a characteristic property
of granular materials, is determined by measuring the steepest angle of the
surface of granular piles. However, the surface of such piles, formed by
pouring granular materials on a horizontal table, is not always the straight line implied by the constant
angle of repose, in the sense that a curved tail is often present at the bottom of the piles.
Alonso and Herrmann~\cite{Herrmann} first investigated the shape of such a curved tail
of a two-dimensional static sandpile.
By introducing the notion of kinks, they built a model to describe the local
slope of the two-dimensional sandpile surface (cf. Eq.~(1) in Ref.~\cite{Herrmann}), and then, using
the observed translational invariance, derived the following expression for the shape of
the tail,
\begin{equation}
x=\frac{h_{m}-h(x)}{\theta_{r}}+l_{e}\ln\frac{h_{m}}{h(x)},	 \label{eq: aH}
\end{equation}
where $\theta_{r}\equiv \tan\theta_{c}$ with $\theta_{c}$ being the angle of repose, $h(x)$ is the height
of the sandpile surface, $x$ represents the corresponding horizontal coordinate, $h_{m}$ is the
maximum height of the sandpile and $l_{e}$$(\equiv l/r)$ is a fitting parameter with $l$ being
the horizontal length of the kink (which typically is of the size of a grain)
and $r$ the corresponding accumulating rate of the grains on the kinks.
The above expression shows that the existence of tails introduces a logarithmic
correction to the straight slope given by the angle of repose, and can be used to
fit the experimental surface shape quite accurately
by adjusting the magnitude of $l_{e}$. However,
the derivation of Eq.~(\ref{eq: aH}) entails three intermediate
parameters, i.e. $\rho$, $l$ and $r$ (where $\rho$ is the density of kinks at height $h$),
but none of them can be determined experimentally(cf. Ref.~\cite{Herrmann} for details), so that
the final fitting parameter $l_{e}$ cannot clearly identify which factors
have an effect on the formation of the curved tail at the bottom of the pile.

In order to answer this query, we study the same problem described above
on the basis of a continuum theory of surface
flows of granular materials proposed by Bouchaud, Cates, Ravi Prakash and Edwards (the BCRE model)~\cite{Edwards}.
This model makes use of two coupled variables to describe the dynamics of the
two-dimensional sandpile surface: the height of the static sandpile, $h(x, t)$, and the
thickness of the rolling grain layer, $R(x, t)$, both of which are a function of
the horizontal coordinate $x$ and the time $t$.
On assuming that the drift velocity, $v$,
of the grains within the rolling layer is independent of $x$, and that diffusion
effects are negligible, the original governing equation of motion for the rolling grains in the BCRE model
can be simplified into
\begin{equation}
  \frac{\partial R(x, t)}{\partial t} =-v\frac{\partial R(x, t)}{\partial x}+ \Gamma(R, h), 	 \label{eq: 1}
\end{equation}
where the interaction term $\Gamma$ accounts for the conversion of immobile grains into rolling grains,
and vice versa. Given that the deviation of the local surface slope $\theta(\equiv-\partial h(x, t)/\partial x)$
of the pile from the critical value $\theta_{r}(\equiv \tan\theta_{c}$ with $\theta_{c}$
being the angle of repose) is everywhere small,
the simplest form of $\Gamma$ is, to a first approximation,
\begin{equation}
\Gamma(h, R) = -\gamma(\theta_{r}-\theta)R(x, t)\equiv -\gamma \left(\theta_{r}+\frac{\partial h(x, t)}{\partial x}\right) R(x, t),	\label{eq: 2}
\end{equation}
where $\gamma$ is a positive constant relative to the property of granular materials,
and can be interpreted as equal to the rate of collision between rolling and static grains.
Moreover, the principle of mass conservation requires that
\[ \frac{\partial h(x, t)}{\partial t}=-\Gamma.\]

In what follows, we shall first directly apply the simplified BCRE model,
i.e. Eqs.~(\ref{eq: 1}) and (\ref{eq: 2}) (henceforth referred to as the BCRE equations),
to investigate the surface shape of two-dimensional piles.

As a granular pile reaches steady growth, the shape of the pile is found to be essentially independent
of time ($t$), i.e. the front of the pile propagates along the $x$ direction with a constant velocity ($u$).
Therefore, we can combine the two independent variables, $x$ and $t$, into one (say $z$)
by means of the following transformation,
\begin{equation}
z=x-ut. \label{eq: tran}
\end{equation}
Consequently, the BCRE Eqs.~(\ref{eq: 1}) and (\ref{eq: 2}) become
\begin{equation}
 -u\frac{ {\mathrm d} R(z)}{ {\mathrm d} z} + v\frac{{\mathrm d} R(z)}{{\mathrm d} z}= u\frac{{\mathrm d} h(z)}{{\mathrm d} z}, \label{eq: 1t}
\end{equation}
and
\begin{equation}
-u\frac{{\mathrm d} h(z)}{{\mathrm d} z} = \gamma R(z) \left(\theta_{r} + \frac{{\mathrm d} h(z)}{{\mathrm d} z}\right).  \label{eq: 2t}
\end{equation}
We can rearrange Eq.~(\ref{eq: 1t}) to yield
\begin{equation}
\frac{{\mathrm d} h(z)}{{\mathrm d} z} = \left(\frac{v-u}{u}\right) \frac{{\mathrm d} R(z)}{{\mathrm d} z},   \label{eq: RH}
\end{equation}
which, when substituted into Eq.~(\ref{eq: 2t}), yields an ordinary
differential equation for the thickness of the rolling grain layer $R(z)$,
\begin{equation}
 (v-u) \left(1 + \frac{\gamma}{u} R(z)\right)\frac{{\mathrm d} R(z)}{{\mathrm d} z} = - \gamma \theta_{r} R(z).  \label{eq: eR}
\end{equation}
We integrate Eq.~(\ref{eq: eR}) and, after applying the boundary condition which determines the flux of pouring grains (i.e. $R=R_{m}$ at $z=0$), obtain
the expression for $R(z)$,
\begin{equation}
z \theta_{r} = \frac{v-u}{\gamma}\ln\frac{R_{m}}{R(z)}+\frac{v-u}{u}\left(R_{m}-R(z)\right),  \label{eq: R1}
\end{equation}
where $R_{m}$ is the maximum thickness of the rolling layer.

Similarly, on integrating Eq.~(\ref{eq: RH}) subject to the the boundary condition
($h=0$ at $R=0$), we find the relationship between $R(z)$ and $h(z)$, i.e.
\begin{equation}
R(z) = \frac{u}{v-u} h(z). \label{eq: RHs}
\end{equation}
Substituting Eq.~(\ref{eq: RHs}) into Eq.~(\ref{eq: R1}), we therefore obtain
the expression for $h(z)$,
\begin{equation}
z = \frac{h_{m}-h(z)}{\theta_{r}} + \frac{v-u}{\theta_{r}\gamma}\ln\frac{h_{m}}{h},   \label{eq: H}
\end{equation}
where $h_{m}$ is the maximum height of the static pile which must satisfy
\begin{equation}
 h_{m} = \frac{v-u}{u} R_{m}.  \label{eq: r1}
\end{equation}

Actually, the maximum thickness of the rolling layer ($R_{m}$) usually
is much smaller than the maximum height of the static pile ($h_{m}$),
which, in view of Eq.~(\ref{eq: r1}), leads to
\[ v \gg 2u. \]
Therefore, Eq.~(\ref{eq: H}) simplifies to
\begin{equation}
z = \frac{h_{m}-h(z)}{\theta_{r}}+\frac{v}{\theta_{r}\gamma}\ln\frac{h_{m}}{h(z)},  \label{eq: Hs1}
\end{equation}
which reduces to,
\begin{equation}
 x = \frac{h_{m}-h(x)}{\theta_{r}} + \frac{v}{\theta_{r}\gamma}\ln\frac{h_{m}}{h(x)},  \label{eq: Hs2}
\end{equation}
provided that the growth of the pile is slow enough to satisfy the quasi-static
assumption, i.e. $ut \sim 0$.

It is interesting to note that Eq.~(\ref{eq: Hs2}) has the same form as Eq.~(\ref{eq: aH})
in that both equations have two terms on the right-hand side: the first one just corresponds to the linear part implied by the angle of repose $\theta_{c}$,
while the second one represents the non-linear part which predicts the existence of the logarithmic tail
at the bottom of the pile. In fact, the only difference between Eqs.~(\ref{eq: aH}) and (\ref{eq: Hs2})
is the coefficient of the logarithmic term: one is $l_{e}$
and the other is $v/\gamma \theta_{r}$.
In contrast to the former (a fitting parameter in Eq.~(\ref{eq: aH})),
the latter consists of an operational parameter ($v$) and two phenomenological constants ($\theta_{r}$ and $\gamma$)
characterizing granular materials, all of which, in principle, can be obtained directly by experimental measurements~\cite{Makes}.
Therefore, according to Eq.~(\ref{eq: Hs2}), we clearly see that, for given $h_{m}$, the length of the curved tail
depends on the rolling velocity ($v$), the angle of repose ($\theta_{c}$) and the rate of collision
between rolling and static grains ($\gamma$).
Moreover, since $v$ as well as $\gamma$ are relevant to the dynamics of the rolling grains on the top of the static pile,
their presence in Eq.~(\ref{eq: Hs2}) throws new light on the
formation of the curved tail at the bottom of the pile. Thus,
for a quasi-static pile, the flow of rolling grains is due to very small perturbations on the pile surface
where the length, time, and velocity are respectively scaled with respect to $a$, $1/\gamma$ and
$a\gamma$ (where $a$ is the size of the grain) so that $v$ must be proportional to
$a\gamma$. However, since $a\gamma$ is essentially independent of the perturbations, so is the value of $v$.
In this case, the formation of the tail is predicted to depend mainly on the size and the shape of granular materials.
On the other hand, if the external disturbances are not so small, the value of $v$ may be relevant to these
disturbances rather than being simply scaled with $a\gamma$. Under this condition, the formation of the tail
depends not only on the properties of the granular materials but also on the operational condition, i.e.
the drift velocity of the grains within the rolling layer. 
 
\begin{figure}
\begin{center}\includegraphics[scale=0.5]{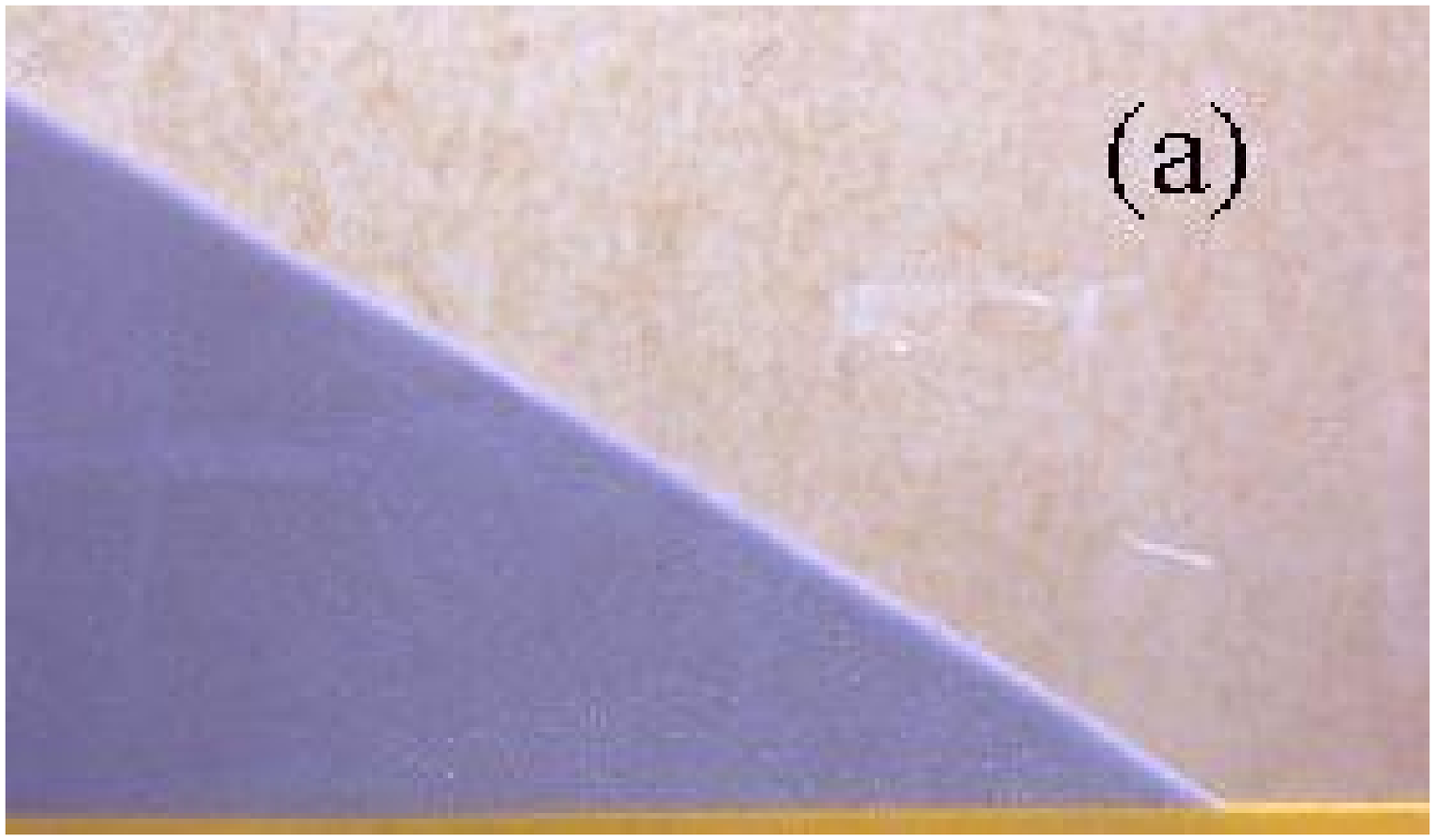}\end{center}
\begin{center}\includegraphics[scale=0.5]{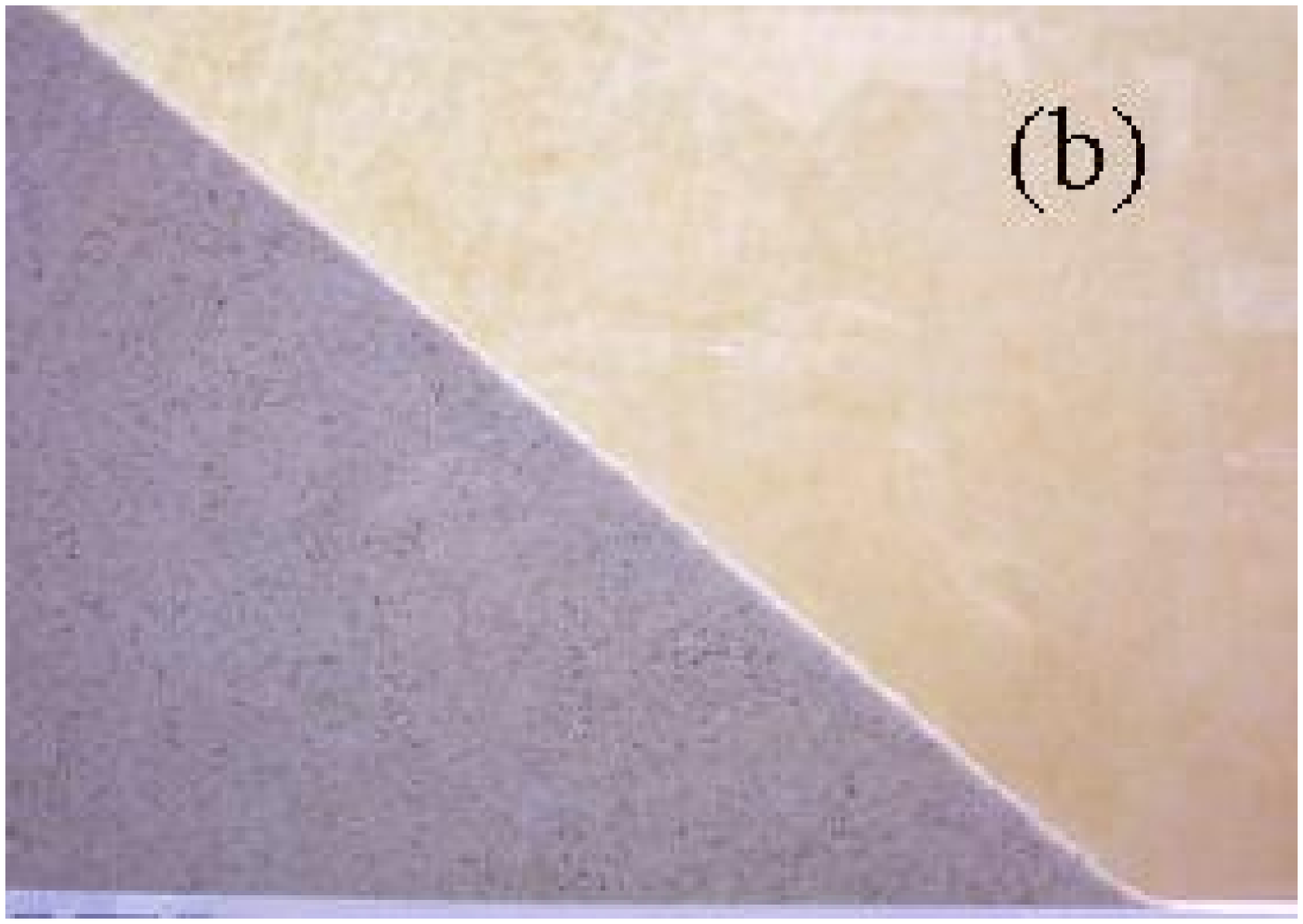}\end{center}
\begin{center}\includegraphics[scale=0.5]{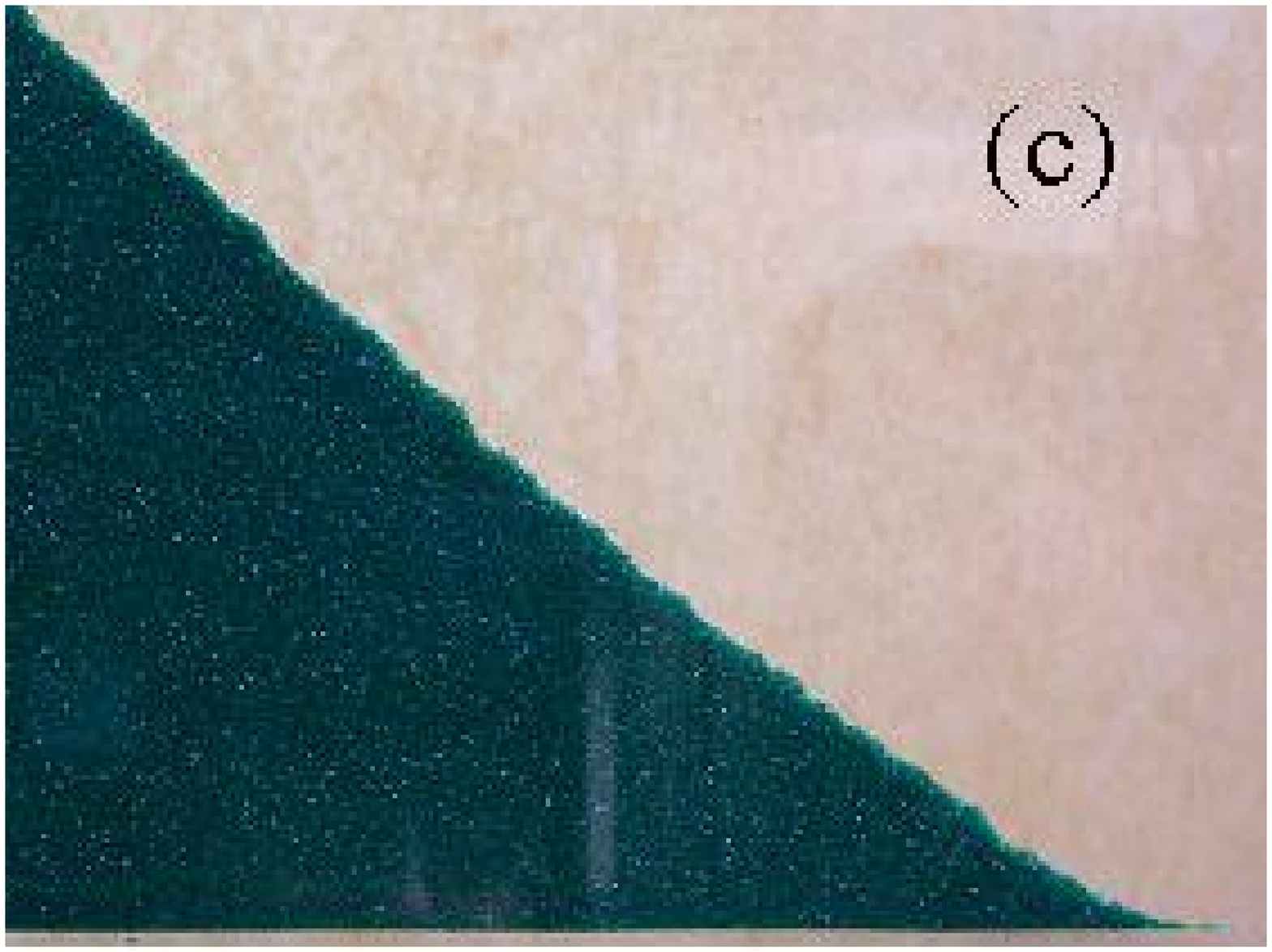}\end{center}
\caption{}
\label{fig1r}
\end{figure}

In order to check the reliability of Eq.~(\ref{eq: Hs2}), we also performed 
experiments with five different granular materials in a rectangular two-dimensional cell made of two vertical 
Plexiglas plates of size $22$cm $\times$ $45$cm with a fixed gap of $2.5$mm.
Granular materials were poured through a funnel placed on the up-left corner of 
the cell. Several pictures of static piles are shown in Fig.~\ref{fig1r}, and
the qualitative comparison between them is listed in Table~1.
For example, no obvious tail exists in the static pile of spherical glass beads (cf. Fig.~1(a) where $a=0.3$mm and $\theta_{c}=30^{\circ}$),
but a long tail is present at the bottom of the static pile of cubic sugar grains (cf. Fig.~1(c) where $a=0.8$mm and $\theta_{c}=45^{\circ}$).
These experimental observations of the static piles show that the formation of the tail is relevant to the properties of granular materials,
which is consistent with the previous discussion following Eq.~(\ref{eq: Hs2}).

\begin{figure}
\begin{center}\includegraphics[scale=0.35]{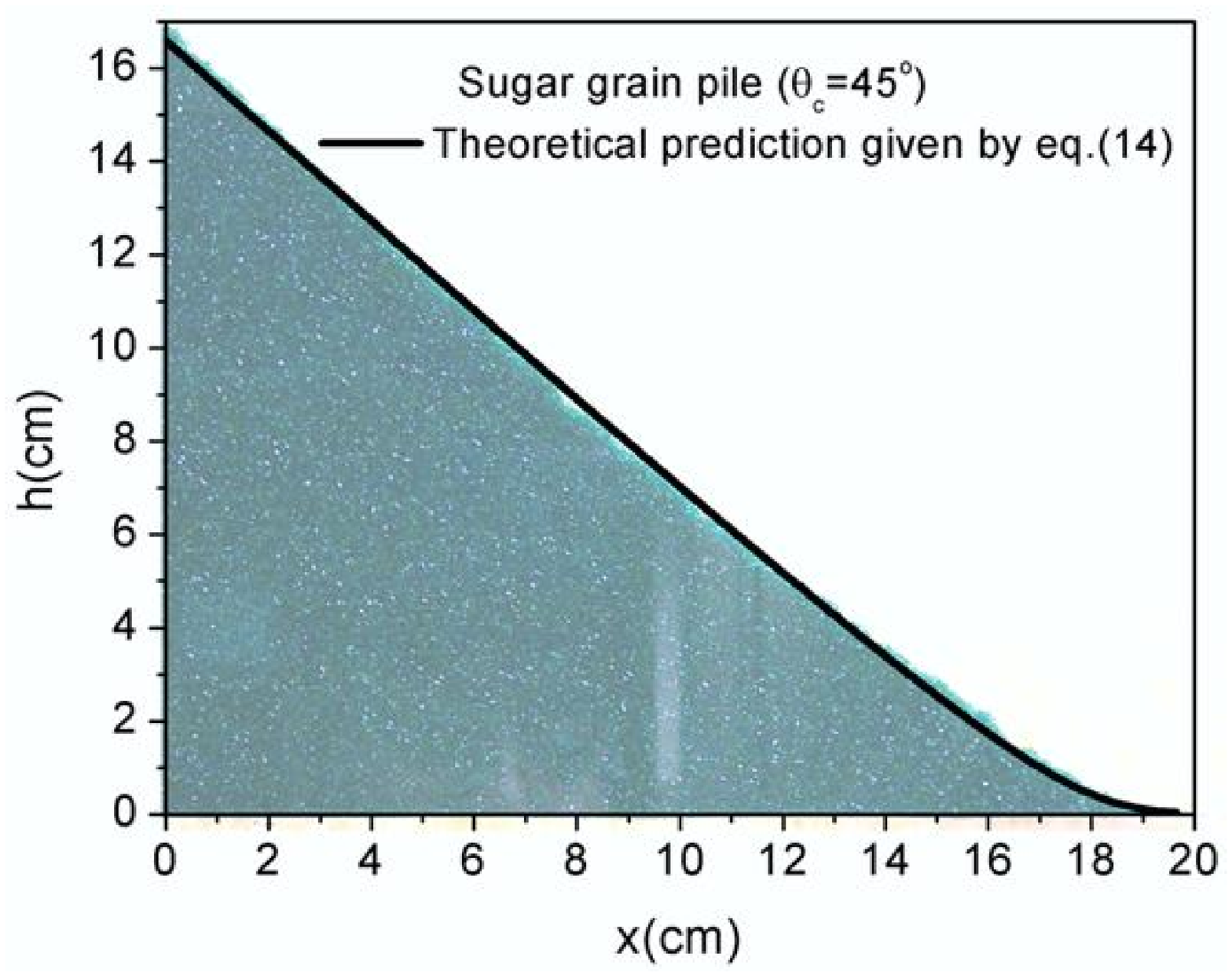}\end{center}
\caption{}
\label{fig2}
\end{figure}

As mentioned earlier, all the parameters in Eq.~(\ref{eq: Hs2}), i.e. $v$, $\theta_{c}$,
$\gamma$ and $h_{m}$, can be obtained by experimental measurements~\cite{Makes}.
However, compared with the last three parameters, it is more difficult to accurately measure $v$
for the case of non-uniform particle velocity profiles within a varying thickness of rolling layer $R(x,t)$.
Under this condition, we can fit the experimental results by taking $v$ to be the only adjustable parameter
and using the values
of $\theta_{c}$, $\gamma$ and $h_{m}$ as determined from the precise experimental measurements. Here, $\theta_{c}$ and $h_{m}$ are
measured in our experiments while $\gamma$ is obtained from early measurements~\cite{Makes}
where the experimental setup as well as granular materials were similar to our case.
In Fig.~\ref{fig2}, a quantitive comparison of the surface shape is made between experiment and theory:
the dark part is the experimental sugar grain pile, and the
solid line represents the surface shape given by Eq.~(\ref{eq: Hs2}).
Clearly, the agreement is satisfactory in both linear and non-linear parts with $v=11.5$cm
and $\gamma=23/$s~\cite{Makes}. The value of $v$ is consistent with previous
estimates \cite{Makes} under the same flux condition as in the present experiments.

\begin{figure}
\begin{center}\includegraphics[scale=0.8]{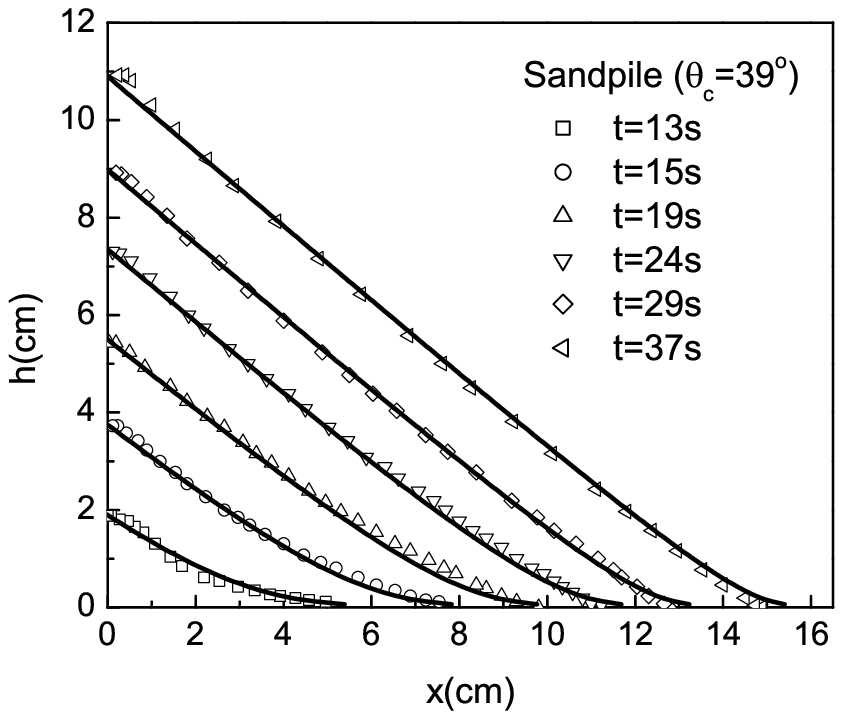}\end{center}
(a)
\begin{center}\includegraphics[scale=0.8]{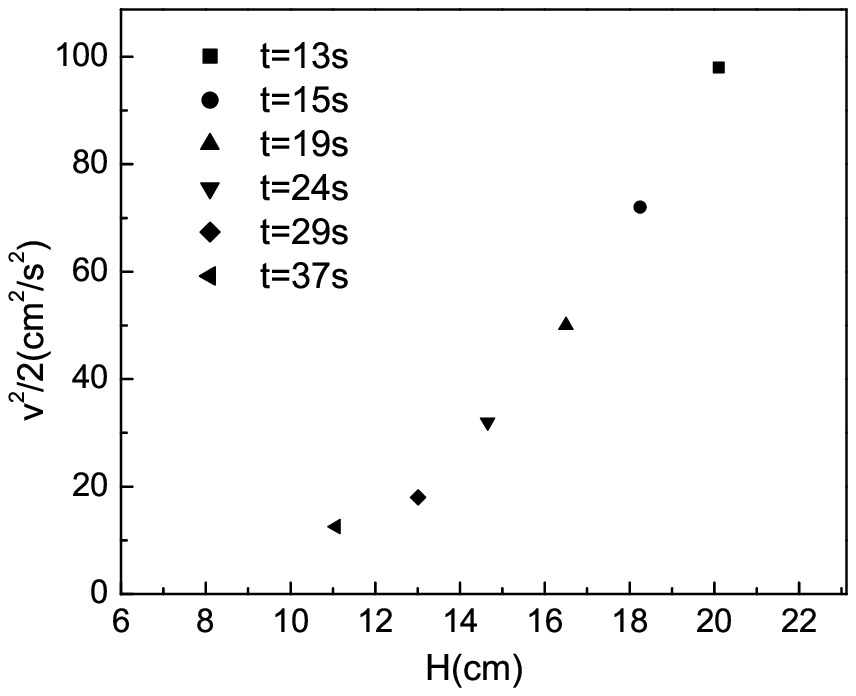}\end{center}
(b)
\caption{}
\label{fig3}
\end{figure}

Next, we investigate the dynamic behavior of pile growth using sand in Fig.~1(b).
During the experiments with sand, we recorded the whole process of the sandpile
growth, and thereby obtained a family of sandpile surface profiles at different growth stages (cf. Fig.~\ref{fig3}(a)).
At the early growth stage (e.g. $t=13s$), the corresponding sandpile
was small and the tail was clearly seen; but when the sandpile grew larger
(e.g. $t=37s$), the corresponding tail became hardly discernible.
Since the position of the funnel was fixed while pouring granular materials into
the cell, the distance between the outlet of the funnel and the top of the sandpile,
called $H$, decreased with the growth of the sandpile. Consequently, the initial velocity
of the rolling particles also decreased with time, which may result in the decrease
of the tail length. To examine this idea, we can still make use of Eq.~(\ref{eq: Hs2})
to fit the experimental profiles of the sandpile surface
at different growth stages by choosing different values of $v$ but
keeping $\gamma$ and $\theta_{r}$ constant.
The fitting results are shown in Fig.~3(a) with solid lines. It is surprising to
see that the theoretical predictions and the experimental data are in good agreement even
in this dynamic case where the shape of the pile is no longer independent of time. Moreover, in light of Fig.~\ref{fig3}(b), 
the value of $v$ chosen in Fig.~\ref{fig3}(a)
is found to be an increasing function of $H$, which is qualitatively consistent with our experimental observation.
In fact, the corresponding value of kinetic energy $v^{2}/2$, as expected by conservation of mechanical energy, is decreasing linearly with a decrease in $H$
for the relative larger $H$ cases and may approach a constant as $H\rightarrow 0 $.
This result implies that the value of $v$ is dominated by the external disturbances from the falling
grains through the distance $H$ if the latter is larger enough; otherwise, it becomes essentially independent of $H$,
and thereby further justifies our choosing different values of $v$ for different
growth stages.

In summary, we have solved the BCRE equations analytically with a simple transformation.
The analytical solution has been used to 
describe the surface shape of the granular pile successfully in both the static
and the dynamic cases. This expression for the surface shape not only predicts
a logarithmic tail as in the early work~\cite{Herrmann}, but also reveals a clear physical picture for its formation.

{\bf Acknowledgments}

The authors would like to thank A. Acrivos and H. Herrmann for discussions. H. Makse
acknowledges support from NSF-DMR-0239504.

\newpage
\begin{table}

\begin{tabular}{|c|c|c|c|c|c|}\hline
Granular  & Shape & Angle of & size & Existence of  & Condition of\\
materials & & repose $\theta_{c}$ & $a$(mm) & curved tail & pile surface \\
\hline\hline
Glass  & Spherical & $30^{\circ}$ & $\cong 0.3$ & No & Smooth\\ \hline
Glass  & Non-Spherical & $39^{\circ}$ & $\cong1.25$ & Yes & Rough\\ \hline
PMMA & Spherical & $37^{\circ}$ & $0.45\sim 0.5$ & Yes & Smooth \\ \hline
Sugar & Cubic & $45^{\circ}$ & $\cong0.8$ & Yes & Rough\\ \hline
Sand & Irregular & $39^{\circ}$ & $\cong 0.3$ & Yes & Smooth\\
\hline
\end{tabular}
\caption[]{{Summary of the experimental results relative to the static granular piles.}
Note that the typical size of the sugar grains is obtained by measuring the volume of 
the cubic grains and calculating the typical size as the cubic root averaged over $10$ 
different grains. The values of the angle of repose are obtained by pouring the grains 
in the cell mentioned earlier and then measuring the resulting angles of piles. 
The cell gap is smaller $(0.25$cm) than that used in Ref.~\cite{Makes}
so that the values of the angle of repose here are greater than those in Ref.~\cite{Makes}
due to the extra friction induced from the vertical walls.}

\end{table}

\clearpage
\newpage
{\large \bf Figure captions}

{\bf Fig 1} Pictures of static granular piles.
(a) Spherical glass bead pile;
(b) Irregular sand grain pile;
(c) Cubic sugar grain pile.

{\bf Fig 2} The surface shape given by Eq.~(\ref{eq: Hs2}) is in good
agreement with the experimental result using sugar grains with $\theta_{c}=45^{\circ}$,
$v=11.5$cm/s and $\gamma=23$/s.

{\bf Fig 3} (a) The surface profiles of the sandpile at different growth stages. The scatters
are the experimental data and the solid lines represent the theoretical results
given by Eq.~(\ref{eq: Hs2}) with $\theta_{c}=39^{\circ}$, $\gamma=20$/s
and different values of $v$. Note that
$t=0$ represents the start point of pouring the sand into the cell.
(b) The value of $v^{2}/2$ with $v$ being the velocity of the gains
within the rolling layer used to fit the pile profiles at different growth stages is decreasing with a decrease in $H$,
the corresponding distance from the outlet of the funnel to the top of the sandpile.
\pagebreak
\clearpage
\pagebreak
\clearpage
\pagebreak
\clearpage

\begin{thebibliography}{99}
\bibitem{Herrmann} J.J. Alonso and H.J. Herrmann, Shape of the Tail of a Two-Dimensional
Sandpile, 1996 {\em Phys. Rev. Lett.} {\bf 76} 4911.
\bibitem{Edwards} J.-P. Bouchaud, M.E. Cates, J.R. Prakash and S.F. Edwards, 
Hysteresis and Metastability in a Continuum Sandpile Model, 1995 {\em Phys. Rev. Lett.}
{\bf 74} 1982.
\bibitem{Makes} H.A. Makse, R.C. Ball, H.E. Stanley and S. Warr, Dynamics of 
Granular Stratification, 1998 {\em Phys. Rev, E} {\bf 58} 3357 .
\end{thebibliography}
\end{document}